\documentstyle [12pt,axodraw] {article}

\catcode`@=12
\begin{document}
\thispagestyle{empty}
\begin{flushright} UCRHEP-T243\\hep-ph/9812344\
\end{flushright}
\vspace{0.5in}
\begin{center}
{\Large \bf Splitting of Three Nearly Mass-Degenerate Neutrinos\\}
\vspace{1.3in}
{\bf Ernest Ma\\}
\vspace{0.3in}
{\sl Department of Physics\\}
{\sl University of California\\}
{\sl Riverside, California 92521\\}
\vspace{1.3in}
\end{center}
\begin{abstract}\ 
Assuming the canonical seesaw mechanism together with an SO(3) family 
symmetry for leptons, broken only by the charged-lepton masses, I show that 
the three neutrinos of Majorana mass $m_0$ are split radiatively in two 
loops by a maximum finite calculable amount of order $10^{-9}~m_0$.  This is 
very suitable for dark matter and vacuum solar neutrino oscillations.  I also 
discuss how atmospheric neutrino oscillations can be incorporated.
\end{abstract}
\newpage
\baselineskip 24pt
There are now a number of experiments \cite{1,2,3} which have varying 
degrees of evidence for neutrino oscillations.  Their implication is that 
neutrinos must have mass, but since only the differences of the squares of 
neutrino masses are relevant in these observations, an intriguing possibility 
exists that each neutrino mass is actually about the same, say of order 1 eV, 
so as to account for part of the dark matter of the universe \cite{4}. 
If so, the theoretical challenge is to understand why neutrinos are nearly 
degenerate in mass \underline {and} why their splittings are so small.

In the context of the canonical seesaw mechanism \cite{5} for small 
Majorana neutrino masses, a common mass $m_0$ of order 1 eV may be obtained 
with the imposition of an SO(3) family symmetry \cite{6}.  Since the 
charged-lepton masses break the above symmetry, the three neutrinos 
$(\nu_e, \nu_\mu, \nu_\tau)$ are then split radiatively in two loops \cite{7} 
by a maximum finite calculable amount of order $10^{-9}~m_0$.  This is a 
consequence of the fact that a Majorana neutrino mass term in the minimal 
standard model comes from an effective operator of dimension five \cite{8,9}. 
Hence vacuum solar neutrino oscillations with $\Delta m^2 \sim 
10^{-10}$ eV$^2$ are natural in this scenario.  With further assumptions, 
a specific model is presented in the following which has maximal 
mixing for solar and atmospheric neutrino oscillations.  
It is also consistent with the absence of neutrinoless 
double beta decay \cite{10}.

Consider three standard-model lepton doublets $(\nu_i, l_i)_L$ and three 
heavy neutrino singlets $N_{iR}$, where the subscript $i$ refers to 
the $(+,0,-)$ components of an SO(3) triplet.  Let $\Phi = (\phi^+, 
\phi^0)$ be the usual Higgs doublet, then the SO(3)-invariant term linking 
$(\nu_i, l_i)_L$ to $N_{iR}$ is
\begin{equation}
f \left[ \left( \bar \nu_+ N_+ + \bar \nu_0 N_0 + \bar \nu_- N_- \right) 
\bar \phi^0 - \left( \bar l_+ N_+ + \bar l_0 N_0 + \bar l_- N_- \right) 
\phi^- \right],
\end{equation}
and the SO(3)-invariant Majorana mass term for $N_{iR}$ is
\begin{equation}
M \left( 2 N_+ N_- - N_0 N_0 \right).
\end{equation}
As $\phi^0$ acquires a nonzero vacuum expectation value $\langle \phi^0 
\rangle = v$, the $6 \times 6$ mass matrix spanning $(\bar \nu_+, \bar \nu_-, 
\bar \nu_0, N_+, N_-, N_0)$ is given by
\begin{equation}
{\cal M}_{\nu,N} = \left[ \begin{array} {c@{\quad}c@{\quad}c@{\quad}c@{\quad}c 
@{\quad}c} 0 & 0 & 0 & m_D & 0 & 0 \\ 0 & 0 & 0 & 0 & m_D & 0 \\ 0 & 0 & 0 & 
0 & 0 & m_D \\ m_D & 0 & 0 & 0 & M & 0 \\ 0 & m_D & 0 & M & 0 & 0 \\ 0 & 0 & 
m_D & 0 & 0 & -M \end{array} \right],
\end{equation}
where $m_D = f v$.  Invoking the well-known seesaw mechanism \cite{5}, the 
$3 \times 3$ mass matrix spanning $(\nu_+, \nu_-, \nu_0)$ is then
\begin{equation}
{\cal M}_\nu = \left[ \begin{array}{c@{\quad}c@{\quad}c} 0 & -m_0 & 0 \\ 
-m_0 & 0 & 0 \\ 0 & 0 & m_0 \end{array} \right],
\end{equation}
where $m_0 = m_D^2/M$.

The physical identites of $\nu_i$ depend on the charged-lepton mass matrix 
which breaks the assumed SO(3) family symmetry.  As a working 
hypothesis, consider the following basis:
\begin{equation}
l_+ = e, ~~~ l_- = c\mu + s\tau, ~~~ {\rm and} ~~~ l_0 = c\tau - s\mu,
\end{equation}
where $c = \cos \theta$ and $s = \sin \theta$. 
The justification for it will come later.  Furthermore, let there be an 
additional mass term $m_1$ for the state $c'\nu_0 + s'(\nu_+ - \nu_-)/
\sqrt 2$, where $c' = \cos \theta'$ and $s' = \sin \theta'$.  This is 
equivalent to breaking the SO(3) symmetry of Eq.~(4) explicitly at tree 
level.  It will be shown 
later where $m_1$ comes from and how it is related to atmospheric neutrino 
oscillations.  The generic statement that the $\nu_i$'s of Eq.~(4) are 
naturally split by a maximum amount of 
order $10^{-9}~m_0$ is independent of the above details.  However, they 
are required for a specific model which explains the present data on both 
solar and atmospheric neutrino oscillations as well as hot dark matter and the 
absence of neutrinoless double beta decay.

In the minimal standard model, any neutrino mass must come from the effective 
operator \cite{8}
\begin{equation}
\Lambda^{-1} \phi^0 \phi^0 \nu_i \nu_j,
\end{equation}
where $\Lambda$ is a large effective mass.   In the canonical seesaw 
mechanism \cite{5}, neutrino masses are generated at tree level \cite{9} and 
may all be different.  However, in the presence of an SO(3) family symmetry 
which is broken only by charged-lepton masses, the radiative splitting is now 
guaranteed to be finite and calculable.  The effective low-energy theory 
is exactly the minimal standard model extended to include a common mass $m_0$ 
for all three neutrinos.  The specific mechanism for their splitting is 
the exchange of two $W$ bosons in two loops \cite{7}, as shown in Fig.~1. 
Since $m_\tau$ is the largest charged-lepton mass by far, the breaking is 
along the $\tau$ direction in lepton space.  (The extra mass term $m_1$ 
breaks the SO(3) symmetry explicitly along a different direction and will 
be considered later.)

The two-loop diagram of Fig.~1 may be evaluated using Eqs.~(3) to (5).  The 
generic structure of the double integral is \cite{7,11}
\begin{equation}
g^4 \int {d^4 p \over (2 \pi)^4} \int {d^4 q \over (2 \pi)^4} 
{1 \over p^2 - m_W^2} {1 \over q^2 - m_W^2} {1 \over p^2 - m^2_{l_i}} 
{1 \over q^2 - m^2_{l_j}} {p \cdot q \over (p + q)^2 - m_0^2} {m_D^2 M \over 
(p + q)^2 - M^2}.
\end{equation}
By dimension analysis, it is clear that the above is proportional to 
$m_D^2/M = m_0$.  Expanding in powers of $m_l^2/m_W^2$, it is also clear 
that there is a universal contribution to $m_0$ which one can disregard, 
and the splitting among the three neutrinos is determined by a term 
proportional to $m_\tau^2/m_W^2$.  (Contributions from $\phi W$ and $\phi 
\phi$ exchanges are negligible because they are at most of order 
$m_\tau^4/m_W^4$.)  Replacing one of the factors involving 
$m_l$ in Eq.~(7), say $(p^2 - m_\tau^2)^{-1}$, with $m_\tau^2/p^4$, the 
resulting integral can be evaluated exactly in the limit $m_0^2 << m_l^2 << 
m_W^2 << M^2$:
\begin{equation}
I = {g^4 \over 256 \pi^4} {m_\tau^2 \over M_W^2} \left( {\pi^2 \over 6} - 
{1 \over 2} \right) m_0 = 3.6 \times 10^{-9}~m_0.
\end{equation}
Consequently, ${\cal M}_\nu$ of Eq.~(4) becomes
\begin{equation}
{\cal M}_\nu = \left[ \begin{array} {c@{\quad}c@{\quad}c} 0 & - m_0 - s^2 I 
& - sc I \\ - m_0 - s^2 I  & 0 & sc I \\ - sc I & sc I & m_0 + 2 c^2 I 
\end{array} 
\right],
\end{equation}
where $m_0$ has been redefined to absorb the universal radiative contribution 
mentioned earlier. 

Whereas the eigenvalues of Eq.~(4) are $-m_0$, $m_0$, and $m_0$, corresponding 
to the eigenstates $(\nu_+ + \nu_-)/\sqrt 2$, $(\nu_+ - \nu_-)/\sqrt 2$, and 
$\nu_0$, those of Eq.~(9) are
\begin{equation}
- m_0 - s^2 I, ~~~ m_0, ~~~ {\rm and} ~~~ m_0 + (1+c^2) I,
\end{equation}
corresponding to the eigenstates
\begin{equation}
{\nu_+ + \nu_- \over \sqrt 2}, ~~~ {c\nu_+ - c\nu_- + s\nu_0 \over \sqrt 
{1 + c^2}}, ~~~ {\rm and} ~~~ {-s\nu_+ + s\nu_- + 2c\nu_0 \over \sqrt 
{2(1 + c^2)}}.
\end{equation}
For positive $m_0$, the eigenvalue $-m_0 -s^2I$ is negative.  However, as 
is well-known, it becomes positive under a $\gamma_5$ rotation of its 
corresponding eigenstate. 
Comparing Eq.~(5) with Eq.~(11) and using Eq.~(10), the probability of 
$\nu_e$ oscillations in vacuum is given by
\begin{eqnarray}
P(\nu_e \to \nu_e) &=&{1+3c^4 \over 2(1+c^2)^2} + {c^2 \over 1+c^2} 
\cos \left( {s^2 \Delta m_0^2 t \over 2E} \right) + 
{s^2 \over 2(1+c^2)} \cos \left( {2c^2 \Delta m_0^2 t \over 2E} \right) 
\nonumber \\ && + 
{s^2c^2 \over (1+c^2)^2} \cos \left( {(1+c^2) \Delta m_0^2 t \over 2E} 
\right),
\end{eqnarray}
where
\begin{equation}
\Delta m_0^2 = 2 m_0 I = 7.2 \times 10^{-9}~m_0^2.
\end{equation}
For $m_0 = 2$ eV and $s^2 = 0.01$, solar neutrino oscillations are then 
interpreted here as mostly $\nu_e \to \nu_\mu$ with 
$\sin^2 2 \theta \simeq 1$ and $\Delta m^2 \simeq 3 \times 10^{-10}$ eV$^2$, 
in good agreement \cite{12} with data \cite{1}. 

The choice of basis given by Eq.~(5) corresponds to the following 
charged-lepton mass matrix linking $(\bar l_+, \bar l_-, \bar l_0)_L$ 
with $(e, \mu, \tau)_R$:
\begin{equation}
{\cal M}_l = \left[ \begin{array} {c@{\quad}c@{\quad}c} m_e & 0 & 0 \\ 
0 & cm_\mu & sm_\tau \\ 0 & -sm_\mu & cm_\tau \end{array} \right].
\end{equation}
This is based on essentially just one assumption, {\it i.e.} that the 
$\nu_e-\nu_e$ entry of the neutrino mass matrix [Eqs.~(4) and (9)] is in fact 
zero.  Neutrinoless double beta decay is then guaranteed to be absent in 
lowest order despite the fact that $m_0$ may be of order 1 eV.  Note that in 
general, one can always choose the $l_{iR}$ basis so that the two zeros 
appear in the first row of ${\cal M}_l$.  After that, one needs to make the 
assumption that $l_+ = e$ to have the two zeros in the first column of 
${\cal M}_l$.  The remaining $2 \times 2$ submatrix is then automatically 
as given.

In addition to ${\cal M}_l$ which breaks the SO(3) family symmetry explicitly, 
consider now the possible origin of $m_1$ for the state $c'\nu_0 + s'(\nu_+ - 
\nu_-)/\sqrt 2$. 
Let there be an extra heavy neutrino singlet $N'$ and an extra Higgs doublet 
$\Phi'$, both of which are odd under a new discrete $Z_2$ symmetry.  In that 
case, the term
\begin{equation}
f' \left[ (c' \bar \nu_0 + s' (\bar \nu_+ - \bar \nu_-)/\sqrt 2) N' 
\bar \phi'^0 - (c' \bar l_0 + s' (\bar l_+ - \bar l_-)/\sqrt 2) N' 
\phi'^- \right] + H.c.
\end{equation}
also breaks the SO(3) family symmetry explicitly and
\begin{equation}
m_1 = {(f' v')^2 \over M'},
\end{equation}
where $M'$ is the Majorana mass of $N'$ and $v' = \langle \phi'^0 \rangle$. 
Now $v'$ may be naturally small compared to $v$ if $\Phi'$ is heavy \cite{13}. 
From the terms $m'^2 \Phi'^\dagger \Phi'$ and $\mu^2 (\Phi'^\dagger \Phi + 
\Phi^\dagger \Phi')$ in the Higgs potential, it can easily be shown that
\begin{equation}
v' \simeq {-\mu^2 v \over m'^2}.
\end{equation}
Since the $\mu^2$ term breaks the discrete $Z_2$ symmetry softly, $v'/v \sim 
10^{-2}$ is a reasonable assumption.  For $M' \sim M$ and $f' \sim f$, a 
value of $m_1/m_0 = 5 \times 10^{-4}$ is thus very natural.  Hence 
atmospheric neutrino oscillations \cite{14} may occur between $\nu_\mu$ and 
$\nu_\tau$ 
with
\begin{equation}
\Delta m^2_{\rm atm} \simeq (m_0 + m_1)^2 - m_0^2 \simeq 2m_0m_1 \simeq 4 
\times 10^{-3} ~{\rm eV}^2
\end{equation}
if $m_0 = 2$ eV, and the mixing angle is $\theta$ if $\theta'$ is small, 
which turns out to be necessary if solar neutrino 
oscillations are to be accommodated at the same time, as shown below.

Inserting $m_1$ into ${\cal M}_\nu$ of Eq.~(9), we find that the mass 
eigenstates are now
\begin{equation}
{\nu_+ + \nu_- \over \sqrt 2}, ~~~ {c'(\nu_+ - \nu_-) \over \sqrt 2} - s' 
\nu_0, ~~~ {s'(\nu_+ - \nu_-) \over \sqrt 2} + c' \nu_0,
\end{equation}
with eigenvalues
\begin{equation}
-m_0 - s^2I, ~~~ m_0 + (c'^2s^2 + 2s'^2c^2 + 2 \sqrt 2 s'c'sc)I, ~~~ 
m_0 + m_1.
\end{equation}
Hence
\begin{eqnarray}
\Delta m^2_{\rm sol} &\simeq& [m_0 + (c'^2s^2 + 2s'^2c^2 + 2 \sqrt 2 
s'c'sc)I]^2 - [m_0 + s^2I]^2 \nonumber \\ &\simeq& 2m_0[2 \sqrt 2 s'c'sc 
+ s'^2(2-3s^2)]I ~\simeq~ 4 \sqrt 2 scs'm_0 I
\end{eqnarray}
if $s' << 1$.  Let $s = c = 1/\sqrt 2$ for maximal mixing in atmospheric 
neutrino oscillations (which is not required by this model, but 
an additional ssumption), then
\begin{equation}
\Delta m^2_{\rm sol} \simeq 4 \times 10^{-10}~{\rm eV}^2
\end{equation}
if $s' = 0.01$ and $m_0 = 2$ eV.

If $m_1$ is absent, then Eq.~(12) governs solar neutrino oscillations, and 
there is no explanation of atmospheric neutrino oscillations.  If $m_1$ is 
present, then atmospheric neutrino oscillations are automatically accounted 
for, but now $s'$ has to be small to explain solar neutrino oscillations.  
Hence $m_1$ should correspond dominantly but not completely to $\nu_0 = 
c\nu_\tau - s\nu_\mu$.  In fact, 
although it is assumed that $\nu_0$ mixes only with $(\nu_+ - \nu_-)/\sqrt 2$, 
the above conclusion will not change if there is also mixing with 
$(\nu_+ + \nu_-)/\sqrt 2$ as long as it is small.

Let the final neutrino mass matrix be rewritten in the basis $(\nu_e, \nu_\mu, 
\nu_\tau)_L$:
\begin{equation}
{\cal M}_\nu = \left[ \begin{array} {c@{\quad}c@{\quad}c} 0 & -c & -s \\ 
-c & s^2 & -sc \\ -s & -sc & c^2 \end{array} \right] ~m_0 ~+~ \left[ 
\begin{array} {c@{\quad}c@{\quad}c} 0 & 0 & -s \\ 0 & 0 & -sc \\ 
-s & -sc & 2c^2 \end{array} \right] ~I ~+ 
\end{equation}

$$\left[ \begin{array} {c@{\quad}c@{\quad}c} s'^2/2 & 
-s'(c's/\sqrt 2 + s'c/2) & s'(c'c/\sqrt 2 - s's/2) \\ -s'(c's/\sqrt 2 + s'c/2) 
& (c's + s'c/\sqrt 2)^2 & -(c'c-s's/\sqrt 2)(c's + s'c/\sqrt 2) \\ 
s'(c'c/\sqrt 2 - s's/2) & -(c'c-s's/\sqrt 2)(c's + s'c/\sqrt 2) & 
(c'c - s's/\sqrt 2)^2 \end{array} \right] ~m_1.$$

The form of the dominant $m_0$ term is exactly the one advocated recently 
\cite{15} if $s=c=1/\sqrt 2$ is assumed.  The transformation matrix between 
$\nu_{e,\mu,\tau}$ and the mass eigenstates $\nu_{1,2,3}$ of Eq.~(19) is 
given by
\begin{equation}
\left( \begin{array} {c} \nu_e \\ \nu_\mu \\ \nu_\tau \end{array} \right) = 
\left( \begin{array} {c@{\quad}c@{\quad}c} 1/\sqrt 2 & c'/\sqrt 2 & s'/\sqrt 2 
\\ c/\sqrt 2 & s's - c'c/\sqrt 2 & -c's - s'c/\sqrt 2 \\ s/\sqrt 2 & 
-s'c -c's/\sqrt 2 & c'c - s's/\sqrt 2 \end{array} \right) \left( \begin{array} 
{c} \nu_1 \\ \nu_2 \\ \nu_3 \end{array} \right).
\end{equation}
In the limit $s'=0$ and $s=c=1/\sqrt 2$, it reduces to
\begin{equation}
\left( \begin{array} {c} \nu_e \\ \nu_\mu \\ \nu_\tau \end{array} \right) = 
\left( \begin{array} {c@{\quad}c@{\quad}c} 1/\sqrt 2 & 1/\sqrt 2 & 0 \\ 
1/2 & -1/2 & -1/\sqrt2 \\ 1/2 & -1/2 & 1/\sqrt 2 \end{array} \right) 
\left( \begin{array} {c} \nu_1 \\ \nu_2 \\ \nu_3 \end{array} \right),
\end{equation}
which shows clearly that both $\nu_e \to \nu_e$ and $\nu_\mu \to \nu_\tau$ 
oscillations are maximal.  Note that the $\nu_e-\nu_e$ entry of Eq.~(23) is 
now $s'^2 m_1/2$, {\it i.e.} of order $10^{-8}$ eV, which is certainly still 
negligible for neutrinoless double beta decay.

In conclusion, the idea of nearly mass-degenerate neutrinos \cite{6,15,16} 
of a few eV should not be overlooked since they may well be the hot dark 
matter of the universe \cite{4}.  A simple and realistic model has been 
proposed, where their splittings are finite calculable radiative 
corrections and are very suitable for vacuum solar neutrino oscillations. 
To allow for atmospheric neutrino oscillations as well, additional explicit 
breaking of the assumed SO(3) family symmetry may be required.  An 
alternative explanation is to have flavor-changing neutrino interactions 
\cite{17}, but that is subject to other serious experimental constraints 
\cite{18}.
\newpage
\begin{center} {ACKNOWLEDGEMENT}
\end{center}

This work was supported in part by the U.~S.~Department of Energy under 
Grant No.~DE-FG03-94ER40837.

\bibliographystyle {unsrt}

\newpage
\begin{center}
\begin{picture}(360,200)(0,0)
\ArrowLine(0,0)(60,0)
\Text(30,-8)[c]{$\nu$}
\ArrowLine(60,0)(120,0)
\Text(90,-8)[c]{$l$}
\ArrowLine(120,0)(180,0)
\Text(150,-8)[c]{$\nu$}
\Text(180,0)[c]{$\times$}
\ArrowLine(240,0)(180,0)
\Text(180,-12)[c]{$N$}
\Text(210,-8)[c]{$\nu$}
\ArrowLine(300,0)(240,0)
\Text(270,-8)[c]{$l$}
\ArrowLine(360,0)(300,0)
\Text(330,-8)[c]{$\nu$}
\PhotonArc(150,-45)(101,26,154)58
\Text(150,68)[c]{$W$}
\PhotonArc(210,45)(101,206,334)58
\Text(210,-68)[c]{$W$}
\end{picture}
\vskip 2.0in
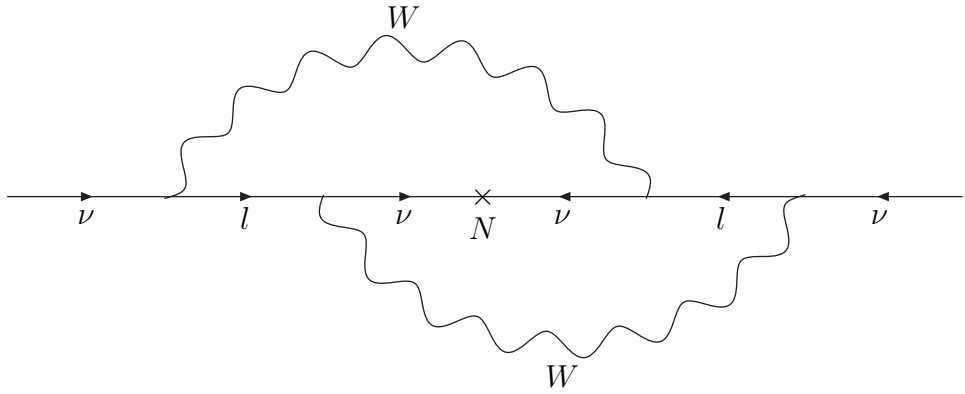
{\bf Fig.~1.} ~ Two-loop radiative breaking of neutrino mass degeneracy.
\end{center}
\end{document}